\documentclass[aps,twocolumn,prresearch
,longbibliography
]{revtex4-1}
\usepackage{amsfonts}
\usepackage{amsmath}
\usepackage{amssymb}
\usepackage{graphicx}
\usepackage{times}
\usepackage[colorlinks,urlcolor=blue,citecolor=blue,linkcolor=blue]{hyperref}

\AtBeginDocument{%
    \newwrite\bibnotes
    \def\bibnotesext{Notes.bib}
    \immediate\openout\bibnotes=\jobname\bibnotesext
    \immediate\write\bibnotes{@CONTROL{REVTEX41Control}}
    \immediate\write\bibnotes{@CONTROL{%
    apsrev41Control,author="08",editor="1",pages="1",title="0",year="1"}}
     \if@filesw
     \immediate\write\@auxout{\string\citation{apsrev41Control}}%
    \fi
}%

\begin{document}

\title{Dynamical quantum phase transitions in a spinor Bose-Einstein
condensate and criticality enhanced quantum sensing}
\author{Lu Zhou$^{1,5,6}$}
\email{lzhou@phy.ecnu.edu.cn}
\author{Jia Kong$^3$}
\author{Zhihao Lan$^4$}
\email{lanzhihao7@gmail.com}
\author{Weiping Zhang$^{2,5,6}$}
\email{wpz@sjtu.edu.cn}
\affiliation{$^1${State Key Laboratory of Precision Spectroscopy, Department of Physics,
School of Physics and Electronic Science, East China Normal University,
Shanghai 200241, China}\\
$^2${School of Physics and Astronomy, and Tsung-Dao Lee institute, Shanghai
Jiao Tong University, Shanghai 200240, China}\\
$^3${Department of Physics, Hangzhou Dianzi University, Hangzhou 310018,
China}\\
$^4${Department of Electronic and Electrical Engineering, University College
London, Torrington Place, London WC1E 7JE, United Kingdom}\\
$^5${Shanghai Branch, Hefei National Laboratory, Shanghai 201315, China}\\
$^6${Collaborative Innovation Center of Extreme Optics, Shanxi University,
Taiyuan, Shanxi 030006, China} }

\begin{abstract}
%the context for your findings 2-3 sentence
Quantum phase transitions universally exist in the ground and excited states
of quantum many-body systems, and they have a close relationship with the
nonequilibrium dynamical phase transitions, which however are challenging to
identify. %the problem 1 sentence
In the system of spin-1 Bose-Einstein condensates, though dynamical phase
transitions with correspondence to equilibrium phase transitions in the
ground state and uppermost excited state have been probed, those taken place
in intermediate excited states remain untouched in experiments thus far. 
%here we show... a summarizing sentence
Here we unravel that both the ground and excited-state quantum phase
transitions in spinor condensates can be diagnosed with dynamical phase
transitions. 
%what did we do and what find out, approach and the main result, specific
A connection between equilibrium phase transitions and nonequilibrium
behaviors of the system is disclosed in terms of the quantum Fisher
information. 
%It is closely related to Loschmidt echoes and imply higher-order
%correlations among eigenstates.
We also demonstrate that near the critical points parameter estimation
beyond standard quantum limit can be implemented. 
%and identify the feasible
%observable.
%what does it mean, the advance over previous work and implications, specific
This work not only advances the exploration of excited-state quantum phase
transitions via a scheme that can immediately be applied to a broad class of
few-mode quantum systems, but also provides new perspective on the
relationship between quantum criticality and quantum enhanced sensing.
\end{abstract}

\maketitle

\affiliation{$^1${State Key Laboratory of Precision Spectroscopy, Department of Physics,
School of Physics and Electronic Science, East China Normal University,
Shanghai 200241, China}\\
$^2${School of Physics and Astronomy, and Tsung-Dao Lee institute, Shanghai
Jiao Tong University, Shanghai 200240, China}\\
$^3${Department of Physics, Hangzhou Dianzi University, Hangzhou 310018,
China}\\
$^4${Department of Electronic and Electrical Engineering, University College
London, Torrington Place, London WC1E 7JE, United Kingdom}\\
$^5${Shanghai Branch, Hefei National Laboratory, Shanghai 201315, China}\\
$^6${Collaborative Innovation Center of Extreme Optics, Shanxi University,
Taiyuan, Shanxi 030006, China} }

\section{Introduction}

% \emph{Introduction.---}
In quantum many-body systems, excited-state quantum phase transitions
(ESQPTs) can be more appealing compared with quantum phase transitions
(QPTs), which refer to quantum criticality aroused in ground states \cite%
{qptbook}. ESQPTs extend the study of criticality to excitation spectra and
have recently been disclosed in several quantum systems \cite%
{esqptreview1,esqptreview2,esqptreview3,esqptcorps2021,lmg1,lmg2}. 
%Typical ESQPTs are accompanied with the appearance of saddle points in the
%phase diagram under classical limit, along which a cumulation of avoided
%crossings are developed in the corresponding quantum spectra of a finite-size
%system.
The criticalities associated with QPTs and ESQPTs can reveal themselves by
nonequilibrium quantum phenomena, especially dynamical phase transition
(DPT) \cite%
{heylPRL2013,heylPRL2015,Heyl_2019,nicolaPRL2021,vajnaPRB2015,jurcevicPRL2017,duPhysRevB2019,fanPhysRevApplied2019,langPhysRevLett2018,flaschnerNP2018,Zhang2017,lmgnature}%
. %\cite{dynamicalqfi,otoclmgesqpt,lmgnature}.

DPT is characteristic of the nonanalyticity in the Loschmidt echo rate
function after quantum quench. More experimentally accessible clue would be
that physical quantities become nonanalytical as a function of time, such as
the order parameter. It is still an open question on the universal
correspondence between DPTs and QPTs, also ESQPTs.

In this work, taking the system of an antiferromagnetic spin-1 Bose-Einstein
condensate (BEC) as an example, we illustrate the relationship between DPTs
and equilibrium phase transitions. Superfluidity and magnetism are
simultaneously achieved in a spinor BEC. Due to the interplay between
intrinsic spin-dependent collision interactions and Zeeman energy splittings
controlled by an external field, the system of a spinor condensate features
a rich phase diagram both in the ground and excited states \cite%
{groundstate2,spinorreview,spinorreview1}. QPTs have been experimentally
explored in the ground state of spin-1 condensates with ferromagnetic \cite%
{baexp} or antiferromagnetic \cite{groundstateexp,dlmexppra,sodiumqpt}
interaction, which show interesting phenomena and applications, such as,
nontrivial dynamics in space \cite{blakiecoarsen,spinorquenchvortex},
Kibble-Zurek mechanism \cite{zurekkz,dlmkz}, preparation of macroscopic
many-body entangled state \cite{youliscience} and surpassing the standard
quantum limit (SQL) \cite{youlipnas}. The authors in \cite{spinoresqpttheory}
showed that the phase transition points can be mapped out through DPT with
measurement on the long-time average of fractional population, which was
used to explore the ESQPT taken place in the uppermost energy level \cite%
{duanlmESQPTexperiment}. However, little efforts have been devoted to the
study of ESQPTs in the intermediate excited states until recently, a
topological order parameter was proposed to characterize ESQPTs in a spinor
BEC \cite{esqpttheory}, whose measurement relies on the precise operation
after one period of spin oscillation and thus can be experimentally
challenging. Besides that, a mimic of ESQPTs in spinor condensates has also
been studied in Raman-dressed spin-orbit coupled BECs \cite%
{celi2021,celi2022}.

Though diverging oscillation periods \cite{microwavefield} and winding
number changing \cite{esqpttheory} are regarded to be linked to ESQPTs, they
can also be explained within mean-field theory and an unambiguous quantum
signature of ESQPTs has not been identified to our knowledge. On the other
hand, only recently has the spin singlet (S) ground state been
experimentally prepared and observed in an antiferromagnetic spinor BEC \cite%
{gerbierfragexp}, since its first prediction in the 90s \cite{pu98}. It is
interesting to explore the DPTs between the S state and other ground states. 
%To resolve
%the singularities in experiments, the most accessible scheme is DPT, i.e.,
%quenching the control parameter across the quantum critical point. The
%dynamics of DPT is determined by the projection of the initial state onto the
%post-quenching eigenstates, and also by the specific features of energy
%spectra and eigenfunctions, thus establishing a connection with ESQPTs
%\cite{brandes2015,santos2015,quench2010,dickeesqpt,quenchprx,dptprx}.
Here, we show that both the QPTs and ESQPTs can be captured with DPT.
Specifically, the nonequilibrium dynamics of DPT could be characterized by
the quantum Fisher information (QFI), which are intimately related to
Loschmidt echoes \cite{dptprx,echoreview,echo1,echo2,tommasoPRA2016}.

% \textcolor{blue}{
Criticality can serve as valuable resources for quantum metrology, finding applications in estimation of external parameters at high sensitivity \cite{zanardiPRA2008,ramsPRX2018,ding2022enhanced,invernizziPRA2008,ivanovPRA2013,tsangPRA2013,parisPRA2014,macieszczakPRA2016,heugelPRL2019,felicettiPRL2020,plenioPRXQuantum2022,Garbe_2022,gietka2022understanding,aybar2022critical,caiprl2021-s}.
Enhanced estimation on the control parameter in a spinor condensate has been studied \cite{spinorsensor1,spinorsensor2},
based on equilibrium phase transitions in the ground states.
It is time-consuming to prepare critical ground states via typical adiabatic evolution, especially in an antiferromagnetic spin-1 condensate due to the closing energy gap between ground and first-excited state \cite{salaPRA2016}.
Motivated by recent study that DPTs can be harnessed for quantum enhanced sensing in a closed quantum system \cite{dynamicalqfi},
we explore the prospect of parameter estimation in a spinor condensate with DPTs. 
% } %Note that 

% \textcolor{blue}{}

%explain the phase diagram including the ground-state spin-singlet phase

\section{QPTs and ESQPTs in an antiferromagnetic spin-1 condensate}

% \emph{QPTs and ESQPTs in an antiferromagnetic spin-1 condensate.---}
We consider a spinor BEC of $N$ atoms with hyperfine spin $F=1$. Within the
single-mode approximation, which enables the internal spin dynamics being
isolated from the external center-of-mass motion, 
%Under the effects of spin-dependent collision interactions and quadratic
%Zeeman energy,
the system is governed by the Hamiltonian ($\hbar=1$) \cite%
{spinorreview,spinorreview1} 
\begin{equation}
\hat{H}=\frac{c}{2N}\hat{S}^{2}-q\hat{N}_{0},  \label{eq:hamiltonian}
\end{equation}
where $c$ and $q$ characterize the inter-spin and effective quadratic Zeeman
energies, respectively. Here, $\hat{S}_{i=x,y,z}=\hat{a}_{\alpha}^{\dagger
}S_{i}^{\alpha\beta}\hat{a}_{\beta}$ are spin-1 vector operators with $\hat {%
a}_{m}$($\hat{a}_{m}^{\dagger}$) the bosonic annihilation (creation)
operators for the magnetic sublevels $m=0$, $\pm1$ and $S_{i=x,y,z}$ the
spin-1 matrices (the indices $\alpha$, $\beta$ are summed over $m$). 
%whereas
%$\hat{Q}_{ij}=\hat{S}_{i}\hat{S}_{j}+\hat{S}_{j}\hat{S}_{i}-\left(
%4/3\right)  \delta_{ij}$ ($\left\{  i,j\right\}  \in\left\{  x,y,z\right\}  $)
%are quadrupole operators.
%should tell N_0 and N
The atom number operators $\hat{N}_{m}=\hat{a}_{m}^{\dagger}\hat{a}_{m}$ and 
$N=\sum_{m}\hat{N}_{m}$. While $q$ can only take positive value if it is
induced by an external magnetic field $B$, i.e., $q\propto B^{2}$, it can be
tuned to both positive and negative values via microwave dressing \cite%
{microwavefield,microwave1,microwave2}. In the following, we will
concentrate on the antiferromagnetic spinor condensate ($c>0$) with zero
magnetization. 
%give the phase diagram %%%%%%%%%%%%%%%%%%%%%%%%%%%%%%%%%%%%%%%
\begin{figure}[!h]
\centering
\includegraphics[width=8 cm]{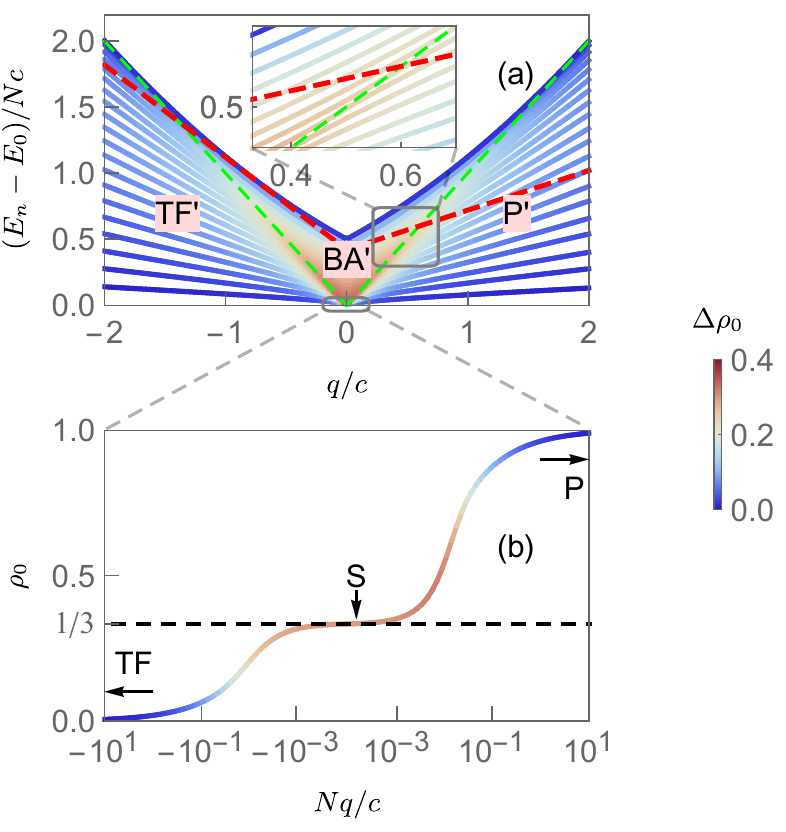} 
\caption{Quantum phases of an antiferromagnetic spin-1 BEC of $N=300$ atoms
with zero magnetization, using the number variance $\Delta\protect\rho_{0}$
in spin-0 component. (a) Excited spectra with every $15$ eigenvalues. The
green-dashed lines are the ESQPT lines. The red-dashed lines represent the
mean-field energy $\mathcal{E}$ of a CSS (\protect\ref{eq_spin coherent})
with $\protect\rho_{0}=0.7$. (b) Ground state varies with $q$ from TF, S to
P.}
\label{fig:phase diagram}
\end{figure}
%%%%%%%%%%%%%%%%%%%%%%%%%%%%%%%%%%%%%%%%%%%%%%%%%%%%%%%%%%%%%%%%
%summarize the phase diagram, can start from the excited state, then goto ground state
%When the atom number $N$ is even,

A sketch of the system phase diagram is given in Fig.~\ref{fig:phase diagram}%
, which is obtained via exact diagonalization of Hamiltonian (\ref%
{eq:hamiltonian}) with an atom number $N=300$. 
%first say ground state then say excited-state
%Due to the competition between the inter-spin collision and the quadratic
%Zeeman term, the spin-1 condensate system favors a rich phase diagram, which
%can be sketched from the eigenspectra shown in
It will be helpful to rewrite Hamiltonian (\ref{eq:hamiltonian}) in a more
generic form as $\hat{H}=\hat{H}_{0}+q\hat{H}_{q}$, and ground state
properties can be recognized as results aroused by the competition between $%
\hat{H}_{0}$ and $\hat{H}_{q}$: 
%Three quantum states are favored in the ground state of the
%system depending on the value of $q$
%\cite{groundstate1,groundstate2,groundstateexp}, which can be better
%understood by rewritting Hamiltonian (\ref{eq:hamiltonian}) in a more generic
%form as $\hat{H}=\hat{H}_{0}+q\hat{H}_{q}$:
(i) For $\left\vert q\right\vert >>c$, the ground state is dominated by $%
\hat{H}_{q}=\hat{N}_{0}$, thus resulting in Polar (P) state and Twin-Fock
(TF) state, with $\rho_{0}\equiv\frac{\left\langle \hat{N}_{0}\right\rangle 
}{N}=1$ and $0$ respectively, in the positive and negative $q$ direction,
associated with vanishing variance $\Delta\rho_{0}\equiv\frac{\Delta \hat{N}%
_{0}}{N}$; (ii) $\hat{H}_{0}\propto\hat{S}^{2}$ restores SO(3) symmetry in a
narrow window of $\left\vert q\right\vert <\frac{c}{N^{2}}$, resulting in S
ground state with $S=0$ for even $N$. %comment on singlet state
State S is massively entangled typical of large variance $\Delta\rho_{0}$,
and with atoms evenly distributed among the magnetic sublevels, representing
a three-fragment mesoscopic quantum state with $\rho_{0}=\frac{1}{3}$. In
the thermodynamical limit of $N\rightarrow\infty$, the S state disappears
and the QPT is characterized by a 1st-order phase transition between the P
and TF state.

%say esqpt
The excited eigenspectra display a cumulation of avoided crossings along $%
E=E_{g}+\left\vert q\right\vert $ (green-dashed lines in Fig.~\ref{fig:phase
diagram}(a), see the inset for a enlarged view), which correspond to
singularities in the density of states under thermodynamical limit. Thus we
refer the lines as the ESQPT lines and the variances $\Delta\rho_{0}$ of the
eigenstates in the vicinity of these lines also achieve maximum values. The
three phases separated by these lines are labled as P$^{\prime}$, TF$%
^{\prime }$ and BA$^{\prime}$ (broken-axisymmetry). While P$^{\prime}$ and TF%
$^{\prime }$ are named according to the corresponding ground states, BA$%
^{\prime}$ is named after the highest energy BA state \cite{baexp,bathe},
which possesses a transverse magnetization perpendicular to the applied
external field and thus breaks the SO(2) axisymmetry.

\section{relation between DPTs and the QFI}

% \emph{DPT and the QFI.---} 
To explore the relation with equilibrium phase transitions, we characterize
DPTs with the QFI, which is defined as the fidelity susceptibility \cite%
{originalqfi,qfianalytical,dynamicalqfi}%
\begin{equation}
F_{Q}\left( q,t\right) =-4\frac{\partial^{2}F\left( q,\delta q,t\right) }{%
\partial\left( \delta q\right) ^{2}}\Big|_{\delta q\rightarrow0},
\label{eq_qfi}
\end{equation}
where the fidelity $F\left( q,\delta q,t\right) \equiv\left\vert
\left\langle \psi\left( q,t\right) |\psi\left( q+\delta q,t\right)
\right\rangle \right\vert =|\langle\psi_{0}|e^{i\hat{H}\left( q\right)
t}e^{-i\hat{H}\left( q+\delta q\right) t}|\psi_{0}\rangle|$ is actually the
Loschmidt echo, and it measures the revival of a state $\left\vert \psi
_{0}\right\rangle $ experiencing time forward propagation under $\hat {H}%
\left( q\right) $ followed by reversed evolution with $\hat{H}\left(
q+\delta q\right) $. One can expect that when the system becomes critical
with $q\rightarrow q_{c}$, the quantum state evolution behaves singularly
and exhibits quite distinct results even for a small $\delta q$, resulting
in prominent decrease of the fidelity and a high $F_{Q}$. An approximate
long-time secular analytic expression for the QFI can be found in Appendix~\ref{app_secular} as 
% \cite{sm}%
\begin{equation}
F_{Q}\left( q,t\right) \simeq4t^{2}\left[ \sum_{n}\left\vert
c_{n}\right\vert ^{2}\left( H_{q}^{n}\right) ^{2}-\left( \sum_{n}\left\vert
c_{n}\right\vert ^{2}H_{q}^{n}\right) ^{2}\right] ,  \label{eq_qfianalytical}
\end{equation}
where $c_{n}=\left\langle \psi_{n}|\psi_{0}\right\rangle $ is the projection
of the initial state $\left\vert \psi_{0}\right\rangle $ onto the
eigenstates $\left\vert \psi_{n}\right\rangle $ of the Hamiltonian (\ref%
{eq:hamiltonian}), and $H_{q}^{n}=\langle\psi_{n}|\hat{H}_{q}|\psi_{n}%
\rangle $. Equation (\ref{eq_qfianalytical}) indicates that a peak in the
QFI can be attributed to either enhanced fluctuations in the order parameter
($H_{q}^{n}=N_{0}^{n}=\langle\psi_{n}|\hat{N}_{0}|\psi_{n}\rangle$), or
those in the overlaps between the initial state and the eigenstates.

\begin{figure}[!h]
\centering
\includegraphics[width=8 cm]{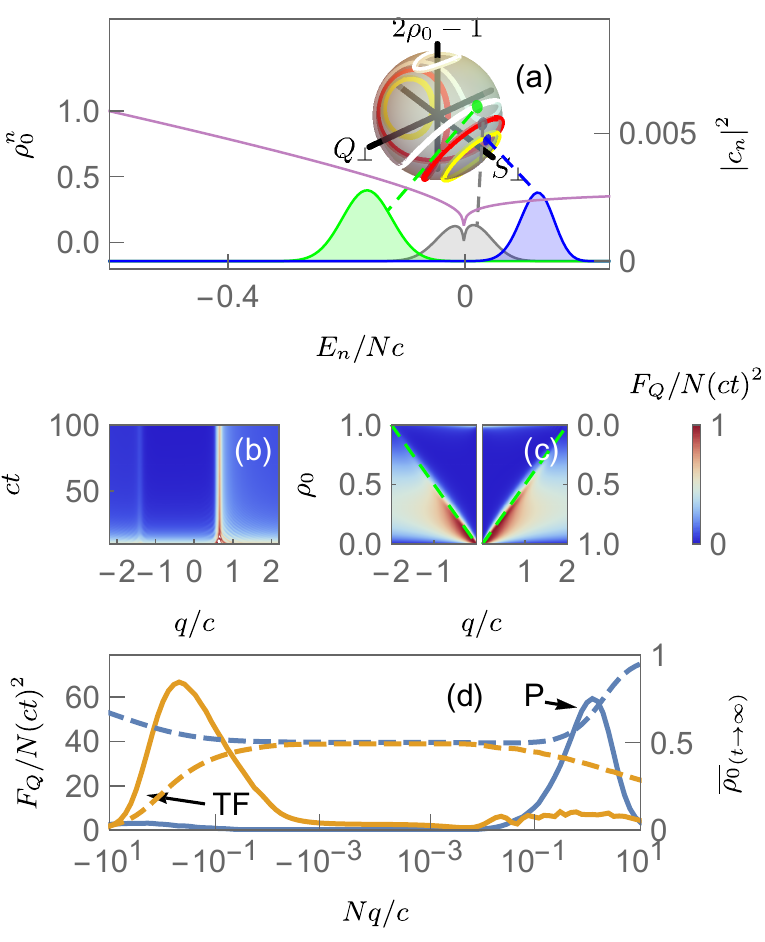} 
\caption{ (a) Slice of the phase diagram Fig.~\protect\ref{fig:phase diagram}
at $q=0.6c$ casted into the $\{S_{\perp}$, $Q_{\perp}$, $2\protect\rho%
_{0}-1\}$ spin-nematic sphere, where the separatrix ($\mathcal{E}=\mathcal{E}%
_{c}$, red line) separates trajectories in the BA$^{\prime}$ phase ($%
\mathcal{E}>\mathcal{E}_{c}$, yellow) from those in the P$^{\prime}$ phase ($%
\mathcal{E}<\mathcal{E}_{c}$, white), and a distribution (the blue, gray,
and green circle) represents a CSS on the sphere with $\protect\rho%
_{0}=0.6,0.7,0.8$ respectively. Eigenstate overlap $\left\vert
c_{n}\right\vert ^{2}$ with these CSSs are plotted (right axis), associated
with the Eigenstate normalized population $\protect\rho_{0}^{n}$ (left axis,
purple curve). (b) Time evolution of the QFI $F_{Q}$ as a function of $q$.
(c) Phase diagram in the $q$-$\protect\rho_{0}$ plane computed with $F_{Q}$
at $ct=10^{3}$. Green-dashed curves refer to ESQPT lines. (d) $F_{Q}$
calculated at $ct=10^{3}$ versus $q$ (left axis) with an initial P state
(blue curve), and an initial TF state (brown curve). The corresponding
dashed curves represent time-averaged population in spin-0 component $%
\overline{\protect\rho_{0}\left( t\right) }|_{t\rightarrow\infty}$ (right
axis). }
\label{fig:qfi}
\end{figure}

To achieve the correspondence between DPTs and equilibrium phase
transitions, one would expect that the overlap between the initial state and
system eigenstates $\left\vert c_{n}\right\vert ^{2}$ has similar singular
distribution as the order parameter around the energy $E_{n}^{c}$ (%
% \textcolor{blue}{
an excited eigenstate will however give zero value of $F_Q$
% }%
). 
% \textcolor{blue}{
Among many possible choices of initial state
% }
, we
propose to use coherent spin state (CSS) $\left\vert \zeta\right\rangle
^{\otimes N}$, with $\left\vert \zeta\right\rangle
\equiv\sum_{m}\zeta_{m}\left\vert m\right\rangle $ and $\zeta=\left( \sqrt{%
\frac{1-\rho _{0}+\rho_{m}}{2}}e^{i\chi_{+}},\sqrt{\rho_{0}},\sqrt{\frac{%
1-\rho_{0}-\rho_{m}}{2}}e^{i\chi_{-}}\right) ^{T}$, where $\rho_{m}=\frac{%
N_{1}-N_{-1}}{N}$, $\chi_{\pm}=\frac{\theta_{s}\pm\theta_{m}}{2}$ with $%
\theta_{s\left( m\right) }$ the spinor phase and magnetization phase
respectively. CSS can be visualized by casting the corresponding mean-field
phase diagram at different $q$ into the spin-nematic phase sphere $%
\{S_{\perp}\text{, }Q_{\perp}\text{, }2\rho_{0}-1\}$ with the transverse
spin $S_{\perp}=\sqrt{\left\langle \hat {S}_{x}\right\rangle
^{2}+\left\langle \hat{S}_{y}\right\rangle ^{2}}$ and transverse
off-diagonal nematic moment $Q_{\perp}=\sqrt{\left\langle \hat {Q}%
_{xz}\right\rangle ^{2}+\left\langle \hat{Q}_{yz}\right\rangle ^{2}}$ \cite%
{spinnematicsphere}, where the quadrapole operators $\hat{Q}_{ij}\equiv\hat{a%
}_{\alpha}^{\dagger}\left[ S_{i}S_{j}+S_{j}S_{i}-\left( 4/3\right)
\delta_{ij}\right] \hat{a}_{\beta}$ (the indices $\alpha$, $\beta$ are
summed over $m$). In the thermodynamical limit, the dynamics of an initial
CSS is characterized by its equal-energy trajectories of the spin-nematic
component on the sphere. As shown in Fig.~\ref{fig:qfi}(a), on the positive $%
q$ side it can be tuned from that in the P$^{\prime}$ phase space (white
line), separatrix dividing the BA$^{\prime}$ and P$^{\prime}$ phase space
(red line linked to the unstable hyperbolic point $n_{0}=0$), and that in
the BA$^{\prime}$ phase space (yellow line). Similar transitions from BA$%
^{\prime }$ to TF$^{\prime}$ phase can take place at $q<0$. The spin
dynamics can be denoted as coherent oscillation with varying amplitude and
period, while for a CSS which is initially localized at the separatrix, it
will become singular with diverging period \cite{zhangPRA2005}.

%figure spin nematic sphere and su3 distribution
%=============================================================
%=============================================================

Taking the CSS with $\left\{ \rho_{0}=0.7\text{, }\rho_{m}=0\text{, }%
\theta_{s}=\theta_{m}=0\right\} $ as an example (whose mean-field energy $%
\mathcal{E}$ is shown as the red-dashed line in Fig.~\ref{fig:phase diagram}%
(a)), at the intersections with the ESQPT line on $q=0.6c$ ($\mathcal{E}=%
\mathcal{E}_{c}=E_{n}^{c}|_{N\rightarrow\infty}$), for a finite system it
represents a distribution on the surface of the spin-nematic sphere with
uncertainty equal to SQL ($1/\sqrt{N}$) and center located on the
separatrix, which is marked as a gray circle in Fig.~\ref{fig:qfi}(a). Since
the mean-field energy of the CSS equals that of the critical saddle point,
it is closer to the eigenstate at which ESQPT takes place as compared with
other higher (blue circle for $\rho_{0}=0.6$) or lower energy CSS (green
circle for $\rho_{0}=0.8$), resulting in the nonanalytical features of $%
\left\vert c_{n}\right\vert ^{2}$ at $E_{n}^{c}$ (gray line), which are not
captured by other CSSs in the P$^{\prime}$ or BA$^{\prime}$ phase (green
line and blue line).

We use the CSS as the initial state to simulate QFI (\ref{eq_qfi}) with atom
number $N=1000$ (see Appendix~\ref{app_numerical} for details about numerical methods) and present the dynamical behavior of $%
F_{Q}/N\left( ct\right) ^{2}$ versus $q$ in Fig. \ref{fig:qfi}(b), where the
normalization with respect to $t^{2}$ is chosen to absorb the expected
long-time growth of $F_{Q}\propto t^{2}$. Around the critical points $%
q_{c}=-1.4c$ and $0.6c$, a prominent increase in the QFI can be observed,
which correspond to the cases where the CSS is centered on the separatrix,
linked to the saddle point $\rho_{0}=1$ and $\rho_{0}=0$, respectively. This
suggests that the quantum dynamics exhibits abrupt change around the
critical points and thus the QFI can serve as an indicator of ESQPTs. These
two QFI peaks in the long-time limit separate the parameter space into three
regions, i.e., the BA$^{\prime}$, TF$^{\prime}$, and P$^{\prime}$ phase,
respectively. Apart from the phase transition region, the QFI displays
damped oscillations.

%\begin{figure}[h]
%\centering
%\includegraphics[width=8 cm]{Fig/qfi.pdf}  \caption{Caption.}%
%\label{fig:qfi}%
%\end{figure}

Motivated by the feasibility that ESQPTs can be distinguished via the QFI,
we map out the excited-state phase diagram by varying the initial CSS. One
simple choice is that keep $\rho _{m}=0$, $\theta _{s}=\theta _{m}=0$ while
vary the value of $\rho _{0}$. For such a state the ESQPT lines display a
monotonic relation with $\rho _{0}$ as $q_{c}=2\left( 1-\rho _{0}\right) $
in the positive $q$ region and $q_{c}=-2\rho _{0}$ in the negative $q$
region. The preparation of such a CSS can be described by the formula%
\begin{equation}
\left\vert \zeta \right\rangle _{\text{initial}}^{\otimes N}=\frac{1}{\sqrt{%
N!}}\left( e^{i\frac{\theta }{2}\hat{Q}_{yz}}\hat{a}_{0}^{\dagger }e^{-i%
\frac{\theta }{2}\hat{Q}_{yz}}\right) ^{N}\left\vert \text{vac}\right\rangle
,  \label{eq_spin coherent}
\end{equation}%
with $\cos \theta =\sqrt{\rho _{0}}$ and $\sin \theta =\sqrt{1-\rho _{0}}$.
In experiments, Eq. (\ref{eq_spin coherent}) corresponds to a process in
which one could first prepare the atoms in the $m=0$ hyperfine state and
then apply a combination of magnetic field ramps and resonant
radio-frequency (rf) pulses \cite{spincoherentstate} to implement polar
state rotation using the quadrapole operator $\hat{Q}_{yz}$. Using the value
of $F_{Q}$ in the long time limit at $ct=10^{3}$, the excited-state phase
diagram is mapped out in the $q$-$\rho _{0}$ plane, as shown in Fig.~\ref%
{fig:qfi}(c). The vertical axis of $\rho _{0}$ is reversed in the right half
($q>0$) with respect to the left half ($q<0$) in order to make a comparison
with the phase diagram in Fig.~\ref{fig:phase diagram}(a). The jump
discontinuities signaling the ESQPTs (green dashed lines) can be well
captured. One can also notice that the QFI in the vicinity of $\left\vert
q\right\vert =2c$ is typically much smaller than that around $q=0$, which
can be traced to the properties of variance $\Delta \rho _{0}$ calculated in
Fig.~\ref{fig:phase diagram}(a).

As for the DPT in ground states, the QFI in the long time limit is
calculated with initial P or TF state respectively, which turns out to
display a peak value at $q\simeq\pm c/N$, as shown in Fig.~\ref{fig:qfi}(d).
These QFI peaks correspond to the QPTs of P$\rightarrow$S and TF$\rightarrow$%
S. For the time-averaged order parameter $\overline{\rho_{0}\left( t\right) }%
|_{t\rightarrow\infty}\equiv\lim_{T\rightarrow\infty}\frac{1}{T}%
\int_{0}^{T}\rho_{0}\left( t\right) dt=\sum_{n}\left\vert c_{n}\right\vert
^{2}N_{0}^{n}/N$, shown as the dashed lines, they donot display any
nonanalyticity for the present small size mesoscopic quantum system. 
%Note that in order to observe the DPTs, which in turn requires high-precision control on the externally applied magnetic or microwave field.

\section{protocol for parameter estimation}

% \emph{$q$-estimation.---}
Despite that in principle the QFI can be measured via performing many-body
quantum state tomography, or measure the excitation rate of a quantum state
upon a periodic drive \cite{goldmanPRB2018,goldmanPRResearch2019,Yu2022}, it
would be complex to implement for a quantum system of hundreds of atoms \cite%
{gerbierfragexp}, and the requirement of real-time measure further prevents
the feasibility of direct derivation of the QFI. In the estimation theory,
the QFI sets the upper bound on the sensitivity of parameter estimation,
i.e., $\Delta q\geq 1/\sqrt{F_{Q}\left( q,t\right) }$, which is termed as
the quantum Cram\'{e}r-Rao bound \cite{originalqfi}. Thus one can get access
to the estimation precision $\left( \Delta q\right) ^{-2}$ through an
observable $\hat{\mathcal{O}}$ as%
\begin{equation}
\left( \Delta q\right) _{\hat{\mathcal{O}}}^{-2}=\frac{\left\vert \partial
_{q}\left\langle \hat{\mathcal{O}}\right\rangle \right\vert ^{2}}{\Delta ^{2}%
\hat{\mathcal{O}}}\leq F_{Q},  \label{eq:snr}
\end{equation}%
with $\Delta ^{2}\hat{\mathcal{O}}=\left\langle \hat{\mathcal{O}}%
^{2}\right\rangle -\left\langle \hat{\mathcal{O}}\right\rangle ^{2}$
representing the variance with respect to the initial state $\left\vert \psi
_{0}\right\rangle $. Eq. (\ref{eq:snr}) indicates that the value of $\left(
\Delta q\right) _{\hat{\mathcal{O}}}^{-2}$ can approach $F_{Q}$ with an
appropriately chosen observable. SQL corresponds to $\left( \Delta q\right)
_{\text{SQL}}^{-2}=Nt^{2}$.

%snr figure
%%%%%%%%%%%%%%%%%%%%%%%%%%%%%%%%%%%%%%%%%%%%%%%%%%%%%%%%%
\begin{figure}[!h]
\centering
\includegraphics[width=8 cm]{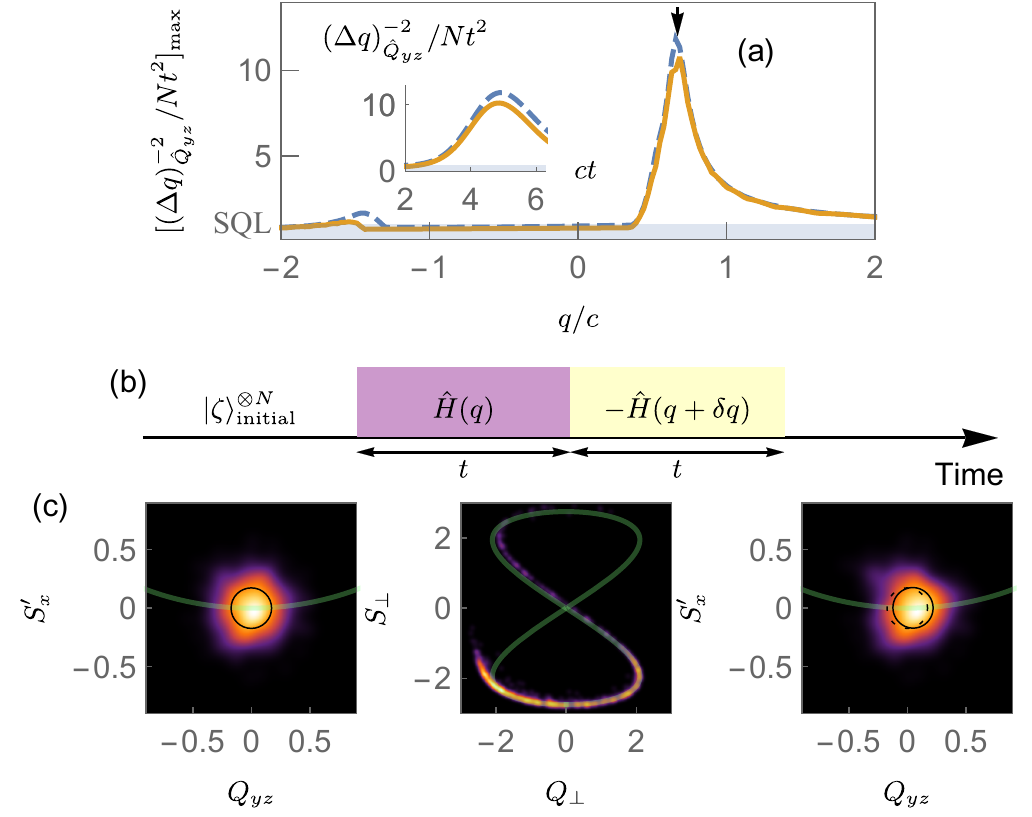} 
\caption{(a) Maximum of the normalized estimation precision $\left( \Delta
q\right) _{\hat{Q}_{yz}}^{-2}$ over time as a function of $q$. The arrow
indicates the precision peak around $q\simeq0.67c$, with the corresponding
time evolution shown in the inset. The dashed lines are those for the QFI,
which set the upper bound of the precision. (b) Schematic showing the
protocol of echo for parameter estimation. (c) Density plot from the
truncated Wigner approximation simulation on the inset of (a), where time is
chosen to be that the optimal precision is achieved, at $ct\simeq4.88$. The
scale is taken to be $100$ atoms and the separatrix is shown in green curve.
Left panel: Distribution of the initial CSS on the $S_{x}^{\prime}$-$Q_{yz}$
space with uncertainty ellipse shown in black curve. Middle panel:
Distribution on the $S_{\perp}$-$Q_{\perp}$ space after the time-forward
propagation. Right panel: Distribution at the end of the echo. The
uncertainty ellipse (black solid curve) shifts in the $Q_{yz}$-direction as
compared with that of the initial state (black dotted curve). All
calculations are for $N=300$. }
\label{fig:snr}
\end{figure}
%%%%%%%%%%%%%%%%%%%%%%%%%%%%%%%%%%%%%%%%%%%%%%%%%%%%%%%%%

Considering that optimal precision is more likely to be achieved for an
observable with small variance, instead of the order parameter $\hat{N}_{0}$%
, we propose to use the quadrapole operator $\hat{Q}_{yz}$ as the observable
(note that $\hat{Q}_{yz}$ is also found to determine the best precision in a
spin-1 condensate interferometry \cite{spinorinterferometry}). While $\hat{S}%
_{x}$ and $\hat{Q}_{yz}$ construct a pair of observables exhibiting
spin-nematic squeezing for an initial P state \cite{chapmanNaturePhysics}, $%
\hat{S}_{x}^{\prime}=\exp\left( i\frac{\theta}{2}\hat{Q}_{yz}\right) \hat{S}%
_{x}\exp\left( -i\frac{\theta}{2}\hat{Q}_{yz}\right) $ and $\hat{Q}_{yz}$
are those for the CSS (\ref{eq_spin coherent}) through a unitary
transformation \cite{uedaPRA2013}. The initial state then constitutes a
minimum uncertainty state for $\hat{S}_{x}^{\prime}$ and $\hat{Q}_{yz}$, as
shown in the left panel of Fig.~\ref{fig:snr}(c). 
% In experiment $Q_{yz}=\left\langle \hat{Q}%
% _{yz}\right\rangle $ can be measured via first exchanging the distribution
% of $Q_{yz}$ with that along $S_{x}$ using microwave pulse, then applying a $%
% \frac{\pi }{2}$ rf rotation about $S_{y}$ and performing Stern-Gerlach
% gradient imaging \cite{chapmanNaturePhysics}. 
Similar to the definition of the QFI in (\ref{eq_qfi}), the precision
estimation also invokes an echo process, which is illustrated in Fig.~\ref%
{fig:snr}(b). We use truncated Wigner approximation to derive the variation
of $Q_{yz}$ after the echo (see Appendix~\ref{app_numerical}), from which the maximum values of $%
\left( \Delta q\right) _{\hat{Q}_{yz}}^{-2}$ are found and they synchronize
well with the behavior of $F_{Q}$, as shown in Fig.~\ref{fig:snr}(a).

% \textcolor{blue}{
In Fig.~\ref{fig:snr}(a)
we take the CSS (\ref{eq_spin coherent}) with $\rho _{0}=0.7$ as
the initial state. A small deviation between the peaks of $\left( \Delta
q\right) _{\hat{Q}_{yz}}^{-2}$ and the mean-field prediction exists, as
indicated by the arrow, where the peak locates around $q\simeq 0.67c$ instead
of the mean-field critical value of $q_{c}=0.6c$, and this can be attributed
to the transient and finite-size effects.
$F_Q$ can be well approximated by $\left( \Delta
q\right) _{\hat{Q}_{yz}}^{-2}$,
and around the critical $q_c$ the QFI scaling beats the SQL sensitivity $\left( \Delta q\right)
_{\text{SQL}}^{-2}\sim N$ (see Appendix~\ref{app_scale} for scaling of the QFI).
These indicate that a parameter estimation precision beyond SQL can be achieved with the onset of criticality.
We have verified that this result remains essentially unchanged when $\rho_0$ is varied for the initial state,
in which the critical points accordingly varies as those have been demonstrated in Fig.~\ref{fig:qfi}(c).
% \textcolor{blue}{
By varying $\rho_0$ from $0$ to $1$,
the corresponding $q_c$ varies in the region of $\left[-2c, 2c\right]$,
with the QFI scaling exponents typically take a value 
% locates in the region of 
$\in\left[1.3, 1.5\right]$.
% }
The precise estimation of the critical $q_{c}$ can then be extended to a much wider
parameter region beyond those at which ground state QPTs take place \cite{spinorsensor1,spinorsensor2}.
% }

To understand the physics beneath the enhanced sensing, we explore the echo
process during which $\left( \Delta q\right) _{\hat{Q}_{yz}}^{-2}$ reaches
its peak value (the inset of Fig.~\ref{fig:snr}(a)). % as
% shown in the left panel of Fig.~\ref{fig:snr}(c). 
After a time-forward evolution under $\hat{H}\left( q\right) $, the atomic
state is dispersed along the separatrix, with its majority surpassing the
saddle point, as shown in the middle panel of Fig.~\ref{fig:snr}(c).
Noticeably that part of the quasi-probability distribution even leaves the
separatrix and enters into the P$^{\prime}$ phase space. This is due to that
the motion near the separatrix is apt to phase space mixing \cite%
{phasespacemixing}. At the end of the echo after experiencing a
time-reversing evolution under $\hat{H}\left( q+\delta q\right) $, the state
approximately recovers the initial CSS (right panel of Fig.~\ref{fig:snr}%
(c)), with a small shift in the $Q_{yz}$ component (compare the uncertainty
ellipse of initial and final states, marked by dotted and solid line
respectively). A small perturbation in the control parameter ($\delta
q=10^{-3}c$ in Fig.~\ref{fig:snr}(c)) can give rise to non-negligible
variation in the observable, and this roots in the sensitive dependence of
quantum state evolution upon the deformation of the separatrix, which is
well captured through an echo process near the critical points. 
%A dynamical QFI can be defined in analogue to its static
%counterpart as \cite{dynamicalqfi}

%To understand this, it is useful to rewrite the Hamiltonian
%(\ref{eq_hamiltonian}) in a more generic form as $\hat{H}=\hat{H}_{0}+q\hat
%{H}_{q}$, from which Note that both these two ingredients play a vital role in
%DPT as a nonanalytic change in the time average of order parameter , which is
%propotional to the second term of the QFI (\ref{eq_qfianalytical}), has been
%used to characterize DPT in the ground and uppermost excited states
%\cite{spinoresqpttheory}. We calculate $\overline{n_{0}\left(  t\right)
%}|_{t\rightarrow\infty}$ as a function of $q$ and $n_{0}\left(  0\right)  $
%and display the resulting quench map in Fig. \ref{fig_qfi}(c), from which one
%can see that along the ESQPT lines from $\left\vert q\right\vert =2c$ to
%$q=0$, $\overline{n_{0}\left(  t\right)  }|_{t\rightarrow\infty}$ behaves more
%singularly. Similar to the QFI, this is aroused by the quantum statistical
%properties of $N_{0}^{n}$ and $c_{n}$, which behave distinctively on the two
%ends of the ESQPT line and reflect different QPT features (for a detailed
%analysis, see \cite{sm}).

\section{conclusion}
% \emph{Conclusion.---}
In summary, we have shown the existence of QPTs and
ESQPTs in an antiferromagnetic spin-1 condensate and demonstrated their
correspondence with DPT, which is characterized using the QFI. We propose
that DPT with the condensate initially prepared in an CSS can be used to
probe the quantum criticality in excited states, which gives rise to a peak
value of the QFI. It can also be used to implement sub-SQL estimation on the
effective quadratic Zeeman energy $q$. 
%Therefore a quantum enhanced sensing on the
%externally applied magnetic field can be achieved in a wide parameter region.
It is interesting to note that the ground-state phase transitions from
symmetry-broken states to the symmetry-restored spin-singlet state can also
be indicated by the DPT. Though we have focused on the system of spinor
condensate, the method of exploring ESQPTs presented here can be applied to
a broad class of few-mode quantum systems.

\begin{acknowledgments}
We thank Han Pu for careful reading on the
manuscript and Keye Zhang for useful discussions. This work is supported by
the Innovation Program for Quantum Science and Technology (2021ZD0303200);
National Key Research and Development Program of China (Grant No.
2016YFA0302001), the National Natural Science Foundation of China (Grant
Nos. 12074120, 11374003, 11654005, 12234014, 12005049, 11935012), the
Shanghai Municipal Science and Technology Major Project (Grant No.
2019SHZDZX01),
Innovation Program of Shanghai Municipal Education Commission (Grant No. 202101070008E00099)
and the Fundamental Research Funds for the Central
Universities. W.Z. acknowledges additional support from the Shanghai Talent
Program. % National Natural Science
% Foundation of China (Grants No. , No. 11774093) and the National Key
% Research and Development Program of China (Grant No. 2016YFA0302001).
L.Z. acknowledges additional support from the Natural Science Foundation of Shanghai (Grant No. 20ZR1418500).
\end{acknowledgments}

\appendix

\section{numerical methods}

\label{app_numerical}

% \textcolor{blue}{
The phase diagram presented in Fig.~\ref{fig:phase diagram} is obtained using the
exact diagonalization method.
% } 
Due to the presence of the SO(2) symmetry in
the Hamiltonian \cite{spinorsymmetry-s}, the generator $\hat{S}_{z}$ is
conserved, i.e., the magnetization $M$ is a conserved quantity. Then the
Hamiltonian matrix $\hat{H}$ written in the $\left\vert
N_{0},M\right\rangle\equiv\left\vert N_{1}=\frac{N-N_{0}+M}{2},N_{0},N_{-1}=%
\frac{N-N_{0}-M}{2}\right\rangle$ basis is block diagonal, for which there
are $2N+1$ blocks with the value of $M$ running from $-N$ to $N$ and each
block has a dimension $\left[ \frac{N-M}{2}+1\right] \times\left[ \frac{N-M}{%
2}+1\right] $ (here $\left[ \cdot\right] $ means taking the integer part).
Each block matrix is tridiagonal and can be diagonalized separately, %
% \textcolor{blue}{
and in Fig.~\ref{fig:phase diagram} a block matrix with $M=0$ is dealt with.
% }

% \textcolor{blue}{
To simulate the quantum Fisher information presented in Fig.~\ref{fig:qfi} we compute the time evolved state $\left\vert
\psi\left( q,t\right) \right\rangle $ with eigenstate expansion.
% } 
The
initial state of the system is described by a coherent spin state $%
\left\vert \zeta \right\rangle ^{\otimes N}$ with (assuming $\rho _{+1}=\rho
_{-1}$)%
\begin{equation}
\zeta =\left( 
\begin{array}{c}
\zeta _{+1} \\ 
\zeta _{0} \\ 
\zeta _{-1}%
\end{array}%
\right) =\left( 
\begin{array}{c}
\sqrt{\frac{1-\rho _{0}}{2}}e^{i\phi _{+1}} \\ 
\sqrt{\rho _{0}}e^{i\phi _{0}} \\ 
\sqrt{\frac{1-\rho _{0}}{2}}e^{i\phi _{-1}}%
\end{array}%
\right) ,  \label{eq_zeta}
\end{equation}%
where equal population in the spin-$\pm 1$ sublevels is assumed. $\left\vert
\zeta \right\rangle ^{\otimes N}$ can be written in the Fock basis as%
\begin{equation}
\left\vert \zeta \right\rangle ^{\otimes N}=\frac{1}{\sqrt{N!}}\left( \zeta
_{+1}\hat{a}_{+1}^{\dagger }+\zeta _{0}\hat{a}_{0}^{\dagger }+\zeta _{-1}%
\hat{a}_{-1}^{\dagger }\right) ^{N}\left\vert 0\right\rangle ,
\label{eq_state full form}
\end{equation}%
which can be expanded in the Fock basis $\left\vert N_{0},M\right\rangle $
as $\left\vert \zeta \right\rangle ^{\otimes N}=\sum_{N_{0},M}f\left(
N_{0},M\right) \left\vert N_{0},M\right\rangle $ with the coefficient%
\begin{align}
f\left( N_{0},M\right) & =\sqrt{\frac{N!}{N_{1}!N_{0}!N_{-1}!}}\left( \sqrt{%
\frac{1-\rho _{0}}{2}}\right) ^{N-N_{0}}\left( \sqrt{\rho _{0}}\right)
^{N_{0}}  \notag \\
& \times \exp \left[ i\left( N_{1}\phi _{+1}+N_{0}\phi _{0}+N_{-1}\phi
_{-1}\right) \right] .  \label{eq_coefficient}
\end{align}

Some spin operators such as $\hat{Q}_{yz}$ couples blocks of different $M$,
which makes its matrix size very large and inconvenient to perform
simulation. We adopt truncated Wigner approximation to study its
dynamics \cite{twa1-s,twa2-s,twa3-s,twa5-s} 
% \textcolor{blue}{
and obtain the
results presented in Fig.~\ref{fig:snr}
% }
. Truncated Wigner approximation states that the Wigner function $W$ for a
quantum state approximately follows the equation%
\begin{equation}
i\hbar\frac{\partial W}{\partial t}\simeq\left\{ H_{W},W\right\} _{C},
\label{eq:twa1}
\end{equation}
where $H_{W}$ is the Wigner-Weyl transform of the Hamiltonian, and $\left\{
\cdots\right\} _{C}$ is the coherent state Poisson bracket. Similarly in the
coherent state picture we treat the operators $\hat{a}_{j}$ ($\hat{a}%
_{j}^{\dagger}$) as complex $c$-numbers $\alpha_{j}$ ($\alpha_{j}^{\ast}$),
and making Wigner-Weyl transform to the Heisenberg equations we have

\begin{equation}
i\hbar \frac{\partial \alpha _{j}}{\partial t}\simeq \left\{ \alpha
_{j},H_{W}\right\} _{C}=\frac{\partial H_{W}}{\partial \alpha _{j}^{\ast }}.
\label{eq:twa2}
\end{equation}%
Truncated Wigner approximation then invokes first sampling the Wigner distribution $W$ with many sets
of $\left\{ \alpha _{j},\alpha _{j}^{\ast }\right\} $, and then for each set
we solve the equation of motion (\ref{eq:twa2}). Any observable of interest
is obtained from the ensemble average. To sample $\left\vert \zeta
\right\rangle ^{\otimes N}$, we first sample the polar state $\frac{1}{\sqrt{%
N!}}\hat{a}_{0}^{\dagger N}\left\vert \text{vac}\right\rangle $ with%
\begin{equation}
\left( 
\begin{array}{c}
\alpha _{1} \\ 
\alpha _{0} \\ 
\alpha _{-1}%
\end{array}%
\right) =\left( 
\begin{array}{c}
\left( a+ib\right) /2 \\ 
\left( e+f\eta \right) e^{i2\pi \xi } \\ 
\left( c+id\right) /2%
\end{array}%
\right) ,  \label{eq:tw3}
\end{equation}%
where $a,b,c,d,\eta $ are independent\ random numbers drawn from Gaussian
distribution with zero mean and unit variance, while $\xi $ is a random
number drawn from uniform distribution in $\left[ 0,1\right] $, and \cite%
{wignersample-s}%
\begin{equation}
e=\frac{1}{2}\sqrt{2N+1+2\sqrt{N^{2}+N}},f=\frac{1}{4e}.  \label{eq:tw4}
\end{equation}%
Unitary transformation to the coherent spin state is equivalent to
performing the rotation%
\begin{equation}
\left( 
\begin{array}{ccc}
\frac{\cos \theta +1}{2} & \frac{\sin \theta }{\sqrt{2}} & \frac{\cos \theta
-1}{2} \\ 
-\frac{\sin \theta }{\sqrt{2}} & \cos \theta & -\frac{\sin \theta }{\sqrt{2}}
\\ 
\frac{\cos \theta -1}{2} & \frac{\sin \theta }{\sqrt{2}} & \frac{\cos \theta
+1}{2}%
\end{array}%
\right) \left( 
\begin{array}{c}
\alpha _{1} \\ 
\alpha _{0} \\ 
\alpha _{-1}%
\end{array}%
\right)  \label{eq:twa5}
\end{equation}%
% \textcolor{blue}{
with $\cos \theta =\sqrt{\rho _{0}}$ and $\sin \theta
=\sqrt{1-\rho _{0}}$
% }
.

We sample a system of $N=300$ with $1000$ trajectories. It has been compared
with the exact quantum mechanical calculations regarding the expectation
values and variances of different spin operators, where good agreements are
found. Truncated Wigner approximation is capable to simulate quantum dynamics in short time scale,
which is enough for us to produce Fig.~\ref{fig:snr} in the main text. However it will
deviate significantly from the exact quantum mechanical calculations when
the evolution time becomes large, due to the omitted high-order terms.

\section{secular approximation of the quantum Fisher information}

\label{app_secular}

%first give the tensor form
The quantum Fisher information (QFI) can be written in a tensor form as \cite%
{qfitensor-s}%
\begin{equation}
F_{Q}\left( q,t\right) =4\left( \left\langle \partial _{q}\psi |\partial
_{q}\psi \right\rangle -\left\vert \left\langle \psi |\partial _{q}\psi
\right\rangle \right\vert ^{2}\right) ,  \label{eq:qfitensor}
\end{equation}%
where $\left\vert \psi \right\rangle =\left\vert \psi \left( q,t\right)
\right\rangle =e^{-i\hat{H}t}\left\vert \psi _{0}\right\rangle $. If $\hat{H}%
=q\hat{H}_{q}$, one can immediately realize that $F_{Q}\left( q,t\right)
=4t^{2}\Delta ^{2}\hat{H}_{q}$ with the variance $\Delta ^{2}\hat{H}%
_{q}=\left\langle \psi _{0}\left\vert \hat{H}_{q}^{2}\right\vert \psi
_{0}\right\rangle -\left\vert \left\langle \psi _{0}\left\vert \hat{H}%
_{q}\right\vert \psi _{0}\right\rangle \right\vert ^{2}$. Recognizing $\hat{H%
}=\hat{H}_{0}+q\hat{H}_{q}$ and $\left[ \hat{H}_{0},\hat{H}_{q}\right] \neq 0
$, we use the identity \cite{deo-s} 
\begin{align}
& \exp \left( i\hat{H}t\right) \frac{\partial }{\partial q}\exp \left( -i%
\hat{H}t\right)   \notag \\
& =-i\int_{0}^{t}dt^{\prime }\exp \left( i\hat{H}t^{\prime }\right) \frac{%
\partial \hat{H}}{\partial q}\exp \left( -i\hat{H}t^{\prime }\right)   \notag
\\
& =-i\mathcal{\hat{H}}_{q},  \label{eq_deo}
\end{align}%
where $\mathcal{\hat{H}}_{q}=\int_{0}^{t}dt^{\prime }\exp \left( i\hat{H}%
t^{\prime }\right) \hat{H}_{q}\exp \left( -i\hat{H}t^{\prime }\right) $ and
this leads to $F_{Q}\left( q,t\right) =4\Delta ^{2}\hat{\mathcal{H}}_{q}$.
Usually, it is not easy to work out the explicit form of $\hat{\mathcal{H}}%
_{q}$ except that the Hamiltonian belongs to some special classes \cite%
{caiprl2021-s}. Using the expansion $\left\vert \psi _{0}\right\rangle
=\sum_{n}c_{n}\left\vert \psi _{n}\right\rangle $, in which $\left\vert \psi
_{n}\right\rangle $ is the eigenstate of $\hat{H}$ with the corresponding
eigenenergy $E_{n}$, we can get 
\begin{align}
F_{Q}\left( q,t\right) & =4t^{2}\left[ \sum_{n}\left\vert
\sum_{m}c_{m}H_{q}^{nm}\text{sinc}\left( \frac{E_{nm}t}{2}\right)
e^{iE_{nm}t/2}\right\vert ^{2}\right.   \notag \\
& \left. -\left\vert \sum_{n,m}c_{n}^{\ast }c_{m}H_{q}^{nm}\text{sinc}\left( 
\frac{E_{nm}t}{2}\right) e^{iE_{nm}t/2}\right\vert ^{2}\right] ,
\label{eq_qfi_expand}
\end{align}%
where $E_{nm}=E_{n}-E_{m}$. In the long-time limit, the sinc function gives
the value of $1$ with $E_{nm}=0$ and zero otherwise. Using an assumption
that only the terms with $E_{nm}=0$ survive in the $t\rightarrow \infty $
limit \cite{dynamicalqfi}\ and considering the fact that the spectrum of $%
\hat{H}$ is nondegenerate, the QFI (\ref{eq_qfi_expand}) can be approximated
by (\ref{eq_qfianalytical}).

We found out that the approximation is quantitatively valid in a moderate
long time. From the viewpoint of phase space mixing \cite{phasespacemixing}%
, a distribution in phase space will eventually mix up in the energy region
it could reach, which will reduce its distinguishability. The
distinguishability of a quantum state with respect to the change of
Hamiltonian parameters, as characterized by the QFI, is also expected to
become coarse in the long-time run. 

\section{scaling of the quantum fisher information}

\label{app_scale}

As standard quantum limit (SQL) correspond to $\Delta q=1/\sqrt{N}$ while in
the Heisenberg limit $\Delta q=1/N$, it is important to know the scaling of
the QFI with system size for understanding whether the dynamical phase
transition corresponding to excited-state quantum phase transition can be
used to implement criticality enhanced sensing beyond the SQL. 
% Dynamical phase
% transitions are intimately related to the initial state and also the
% properties of the associated phase transitions, 
The properties of initial state and associated phase transitions play a
central role in dynamical phase transitions, thus also affect the QFI
scaling. Here we illustrate with the example studied in the main text, i.e.,
the system is prepared in an initial coherent spin state with $%
\left\{\rho_0=0.7, \rho_m=0, \theta_s=\theta_m=0\right\}$. With this initial
state, the dynamical phase transitions corresponding to excited state
quantum phase transitions are expected to take place at $q_c=-1.4c$ and $%
0.6c $ in the $N\rightarrow\infty$ limit.

% figure for qfi scaling
\begin{figure}[!h]
\centering
\includegraphics[width=7.5 cm]{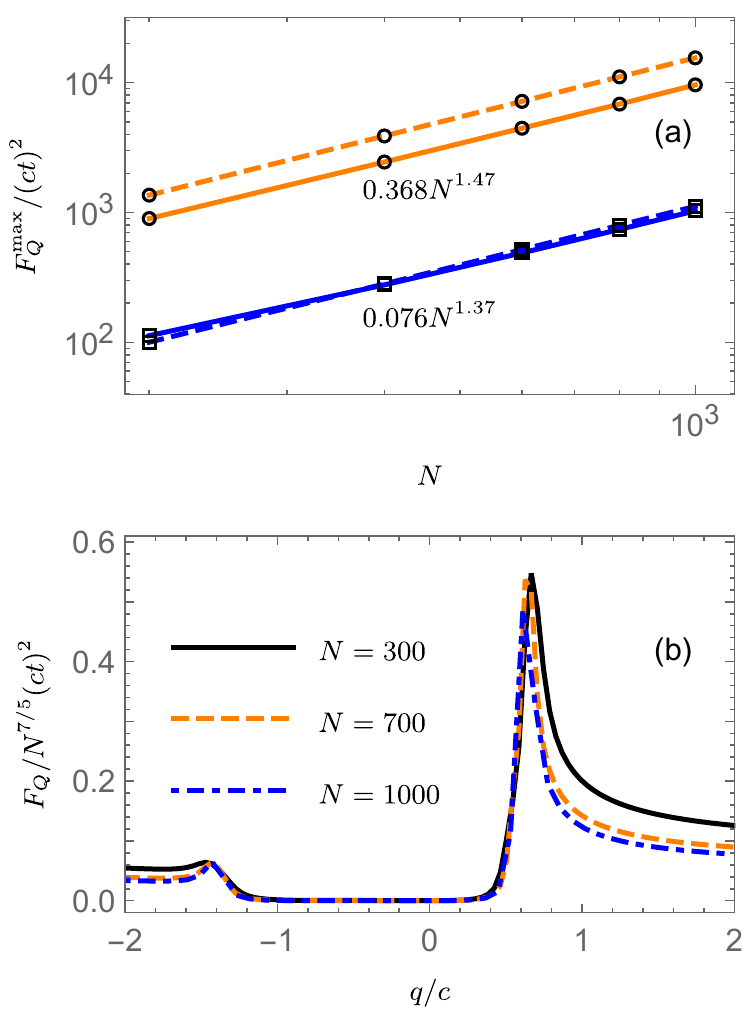}
\caption{Scaling of the QFI. (a) $F_Q^{\text{max}}$ versus $N$ around $%
q_c=-1.4c$ (square) and $q_c=0.6c$ (circle). The numerical and analytical
results are represented by those linked by solid and dashed lines,
respectively. (b) $F_Q$ scaled by $N^{7/5}$ as a function of $q$. }
\label{fig:scale}
\end{figure}

We then study the QFI scaling around these two critical points. The results
are demonstrated in Fig.~\ref{fig:scale}(a), in which the maximum QFI (dots
linked solid lines) are numerically derived around critical points of $%
q_c=-1.4c$ (square) and $q_c=0.6c$ (circle) respectively in the long time
limit ($ct=1000$). Compared with the long-time analytical results predicted
by (\ref{eq_qfianalytical}) (dots linked by dashed lines), we found that the
scaling around the negative $q_c$ are almost identical. On the other hand,
while the analytical QFI is slightly larger than the numerical results
around the positive $q_c$, their scaling behavior (slope of the line)
coincide. The differences between analytical and numerical results are due
to the coarse of quantum state distinguishability in the long-time run, as
discussed in the end of Appendix~\ref{app_secular}.

From data fitting on numerical results we extract the scaling exponent of
the QFI, which give the value of $1.37$ and $1.47$ respectively, for the
dynamical phase transitions on the negative and positive $q$ side. That the
scaling exponent larger than $1$ indicate the feasibility of achieving a $q$%
-estimation precision beyond SQL as $\left( \Delta q\right) ^{-2}\sim
F_{Q}>\left( \Delta q\right) _{\text{SQL}}^{-2}$. In Fig.~\ref{fig:scale}(b)
we demonstrate the QFI scaling as a function of $q$. 
% and found that $7/5$ to be a good approximate overall scaling exponent.
Noticeably that peak of the QFI shifts to a value larger than the positive $%
q_c$ for small system size $N$, this is due to the finite size effect which
have also been illustrated in Fig.~\ref{fig:snr}.

\section{experimental consideration}

The echo including first evolves an initial state $\left\vert \psi
_{0}\right\rangle $ forward with $\hat{H}\left( q\right) $ and then backward
with $\hat{H}\left( q+\delta q\right) $. To implement this, one needs to
reverse the sign of the Hamiltonian $\hat{H}=\hat{H}_{0}+q\hat{H}_{q}$ such
that the system can experience time-reversing evolution. In experiments, the
sign of the control parameter $q$ can be varied via microwave dressing \cite{qsignreverse,microwave1,microwave2}. On the other hand, the sign of $%
\hat{H}_{0}$ is determined by the spin-dependent interaction coefficient $c$%
, which may be reversed via transferring the atoms from the $F=1$ hyperfine
manifold to that of $F=2$ \cite{oberthalerPRL2019}. Another promising
technique could be exploited is the photon-mediated spin-exchange
interactions, which was experimentally realized recently with the aid of a
cavity light field \cite{signreverse,lmgscience,massonPRL2017}. Other methods capable
of manipulating spin-dependent collision interactions, such as
photoassociation \cite{jinghuiphotoassociation}, also exist.

% \bibliographystyle{apsrev4-1}
% \bibliography{main}
%merlin.mbs apsrev4-1.bst 2010-07-25 4.21a (PWD, AO, DPC) hacked
%Control: key (0)
%Control: author (8) initials jnrlst
%Control: editor formatted (1) identically to author
%Control: production of article title (0) allowed
%Control: page (1) range
%Control: year (1) truncated
%Control: production of eprint (0) enabled
%
 
\end{document}